\documentclass[preprint,12pt]{elsarticle}
\usepackage{etoolbox}
\makeatletter
\def\ps@pprintTitle{%
	\let\@oddhead\@empty
	\let\@evenhead\@empty
	\def\@oddfoot{\reset@font\hfil\thepage\hfil}
	\let\@evenfoot\@oddfoot
}
\makeatother
\counterwithin{figure}{section}
\counterwithin{table}{section}
\usepackage{graphics}
\usepackage{amsmath, amsthm, amssymb}
\usepackage{natbib}
\usepackage[colorlinks=false]{hyperref}
\usepackage{lscape}
\usepackage[utf8]{inputenc} 
\usepackage{amsthm}

\usepackage{enumerate}
\usepackage{mathrsfs}
\usepackage{setspace}
\usepackage{multirow}
\usepackage{graphicx}
\usepackage{amsfonts}
\usepackage{verbatim}
\usepackage{amssymb}
\usepackage{amsthm}
\usepackage{caption}
\usepackage{subcaption}
\usepackage{multirow,bigstrut,bigdelim}

\theoremstyle{plain}

\theoremstyle{remark}

\numberwithin{equation}{section}
\usepackage[mathscr]{euscript}
\usepackage{geometry} 
\usepackage{graphicx,lscape} 
\usepackage{booktabs}
\usepackage{array}
\usepackage{paralist} 
\usepackage{verbatim}
\makeatletter
\renewcommand{\fnum@figure}{\textbf{Fig. \thefigure}}
\makeatother
\usepackage{caption}
\usepackage[labelsep=space]{caption}
\biboptions{comma,round,authoryear}

\begin{document}
	
	\begin{frontmatter}

        \title{ Regression analysis of cure rate models with competing risks subjected to interval censoring  }

            \author{Silpa K.}
		\author{Sreedevi E. P.$^ \dag$ } 
		\author{P. G. Sankaran}
		\address{Department of Statistics, Cochin University of Science and Technology, Cochin 682 022, Kerala, India\\ Corresponding author email: sreedeviep@gmail.com  }

				\begin{abstract}
   In this work, we present two defective regression models for the analysis of interval-censored competing risk data in the presence of cured individuals, viz., defective Gompertz and defective inverse Gaussian regression models. The proposed models enable us to estimate the cure fraction directly from the model. Simultaneously, we estimate the regression parameters corresponding to each cause of failure using the method of maximum likelihood. The finite sample behaviour of the proposed models is evaluated through Monte Carlo simulation studies. We illustrate the practical applicability of the models using a real-life data set on HIV patients.
			\end{abstract}
		
\begin{keyword}
Cure fraction, Interval censoring, Competing risks, Defective distribution, Gompertz model, Inverse Gaussian model, Maximum likelihood estimation.
\end{keyword}	
  
  \end{frontmatter}
\section{Introduction}

Interval censoring occurs in various practical settings, particularly in medical studies where the subjects are periodically assessed for disease recurrence or progression. Such data are commonly observed in follow-up clinical studies, like those involving cancer biochemical recurrence or AIDS drug resistance, where the event of interest has low fatality rates and patients are monitored at regular intervals (\cite{sun2006statistical}). In interval censoring, the time to an event of interest is not directly observed and is known to fall within a specific time interval.
In survival studies, competing risks are often observed, where each subject may experience an event or failure due to one of several possible causes. 
 Competing risks data are typically analyzed using one of the following three formulations: 
 \begin{itemize}
     \item The cause specific hazard function (CSHF), $h_j(t)$, given by
     \begin{equation}\label{hf}
       h_j(t)=\lim_{\Delta t\rightarrow 0}\frac{P(t \leq T < t+\Delta t, J=j|T\ge t)}{\Delta t}, ~~~~~~j=1,2, \dots , k.
    \end{equation}
    \item The cumulative incidence function (CIF) for events of type $j$ given by
    \begin{equation}
       F_j(t)=P(T \leq t, J=j),\quad j=1,2,\dots,k.
    \end{equation}
    \item 
    The cause specific sub-survival function given by 
    \begin{equation} 
    S_{j}(t) = \exp\left\{ -\int_{0}^{t}h_{j}(t)dt\right\},~~~ j=1,2,\ldots ,k,
    \end{equation} 
    where $h_{j}(t)$ is the CSHF.
\end{itemize}
Sub-survival functions are used in literature by several authors, including \cite{carriere2000comparing} and \cite{sreedevi2017semi}. One could refer to \cite{porta2008role} for a detailed study on sub-survival functions. 

In real life studies on lifetime data, it is common to observe interval censored failure times that are exposed to competing risks. Several authors considered the analysis of interval censored competing risks data in the literature. \cite{hudgens2001nonparametric} derived the nonparametric estimation methods for interval-censored competing risks data, while \cite{jeong2006direct} and \cite{hudgens2014parametric} studied parametric modeling for interval censored competing risks data.
Some recent important works in this area include \cite{mao2017semiparametric}, \cite{park2019semiparametric}, and \cite{mitra2020analysis}, among others.

On many occasions in survival studies,  the event of interest never occurs for some individuals, despite the sufficient follow-up time. These individuals are considered to be cured (immune) individuals. Cure models are used to analyze survival data in the presence of cured individuals. The most popular method for analyzing cure rate data is the mixture cure rate model (MCM) proposed by \cite{boag1949maximum} and studied by \cite{berkson1952survival}. In the mixture model approach, a specific parameter accounts for the cure rate, with the survival function given by $S(t)=p+(1-p)S_{0}(t)$, where $p$ is the cure fraction and $S_{0}(t)$ is the survival function of the uncured individuals. Several authors, including \cite{farewell1986mixture}, \cite{peng2000nonparametric}, \cite{sy2000estimation},
 and \cite{zhang2007new}, have studied the standard mixture model. \cite{maller1996survival} and \cite{ibrahim2001bayesian} provide excellent references on this topic.
 
Studies on cure rate models with interval censoring are available in the literature including,  \cite{kim2008cure}, \cite{xiang2011mixture}, and \cite{shao2014semiparametric}. Similarly, the analysis of competing risks data with a cure fraction has also been explored. The studies on various approaches for modeling competing risks data with a cure fraction include \cite{choi2015efficient}, \cite{choi2018semiparametric}, \cite{nicolaie2019vertical}, and references therein. 

 Defective models and promotional time cure models serve as efficient alternatives to mixture cure models. 
 The promotion time cure model, proposed by \cite{tsodikov1996stochastic}, is characterized by an improper survival function $S(t)=e^{-\theta F(t)}$, where $F(t)$ is an unknown distribution of a non-negative random variable and $\theta > 0$. In promotion time cure models, the cure probability is obtained by $\lim_{t \to \infty} S(t)= e^{-\theta}>0$.
 Defective models introduced by \cite{balka2009review}  simplify the modeling process by enabling the direct estimation of the cure rate using an improper distribution. The standard distributions are modified by changing the domain of parameters to obtain defective models. In defective models, the cure probability is estimated as,  $\lim_{t \to \infty} S(t)=p,$ with $p \in (0, 1).$  One major advantage of defective models over standard mixture models is that they require one fewer parameter to be estimated in the likelihood function. Commonly used defective distributions in the literature include the defective Gompertz distribution (\cite{gieser1998modelling}), the defective inverse Gaussian distribution (\cite{balka2011bayesian}), and the defective distributions based on the Kumaraswamy and the Marshall-Olkin families (\cite{rocha2017new}). Besides, defective models have been studied by some other authors, including  \cite{martinez2018new}, and \cite{peres2022bivariate}, among others. \cite{calsavara2019defective} proposed defective regression models for modeling interval-censored data.


The scenario where a proportion of lifetimes exposed to competing causes of failure is interval censored, while the remaining proportion of the population does not experience any events during the entire follow up period, gives rise to interval censored competing risks data in the presence of cured individuals.  For example, consider a study on death and loss to HIV care in sub-Saharan Africa. The data consists of 3000 patients who initiate antiretroviral treatment (ART). For each patient, the event is known to occur between two successive intervals, with two competing causes: either loss to HIV care or death. Of the 3000 patients, 2154 do not experience any events during the follow-up period, strongly indicating the presence of cured individuals in the population. The HIV data is studied in detail in Section 4. The real life situation discussed above, make it necessary to model interval censored competing risks data in the presence of cured individuals.


Although interval censored data in the presence of cured individuals, as well as competing risks models in the presence of cured individuals, have been explored separately, cure models accommodating interval censored competing risks data have not been studied yet. Motivated by this, in this study, we propose two defective regression models for analyzing interval-censored competing risk data in the presence of cured individuals. We present two defective regression models: the Gompertz defective model and the inverse Gaussian defective model.




The rest of the article is organized as follows: Section 2 introduces the Gompertz defective model and the inverse Gaussian defective model, and outlines the inference procedures for an interval-censored competing risks setup. In addition, in Section 2, we describe the maximum likelihood estimation of regression parameters and the newly developed estimator for the overall cure fraction. Section 3 reports the results of simulation studies assessing the finite sample performance. In Section 4, we demonstrate the practical applicability of the models using data from an HIV study on death and loss to HIV care in sub-Saharan Africa. Finally, Section 6 provides the concluding remarks with a discussion on future works.
\section{Inference Procedures}\label{sec1}
\subsection{Method}
Consider a study involving $n$ individuals, where $J$ $\in$ $\{1,2,...,k\}$ represents the cause of failure for an individual who can experience one of $k$ competing events. Let $T$ denote the lifetime, which is known to have occurred within a specific interval. Let $(U_{i},V_{i}]$ represent the observed interval, where $U_{i}$ is the last inspection time before the event and $V_{i}$ is the first inspection time just after the event, such that $U_{i} < V_{i}.$ We observe a $p \times 1$ vector of covariates $X$ corresponding to each individual. 
We define the indicator function $\delta_{ij}$ by
    \begin{equation*}
    \delta_{ij} = \left\{
    \begin{aligned}
        & 1 && \text{if the } i \text{-th individual experiences the } j \text{-th cause of failure} \\
        &    && \text{during the interval } (U_{i}, V_{i}], \\
        & 0 && \text{otherwise},
    \end{aligned}
    \right.
\end{equation*}
    for $j=1,2, \ldots ,k$ and $i=1,2, \ldots ,n$.
Let $\delta_{i} =1-\sum_{i=1}^{n}\sum_{j=1}^{k}\delta _{ij}$, and for right censored observations, $\delta_{i} =1$. 
The likelihood function for interval censored competing risks data is expressed as
\begin{equation}
    L(\vartheta )=\prod_{i=1}^{n}\left \{\prod_{j=1}^{k}\left \{ S_{j}(U_{i},\vartheta_{j})-S_{j}(V_{i},\vartheta_{j}) \right \}^{\delta _{ij}}\left \{\prod_{j=1}^{k} S_{j}(U_{i},\vartheta_{j}) \right \}^{1-\delta_{i}}\right \}.
\end{equation}
Then the log-likelihood function is given by
\begin{equation}
    l(\vartheta)=\log L(\vartheta),
\end{equation}
where $\vartheta$ is the vector consisting of elements of $\vartheta_{1}, \vartheta_{2}, \ldots ,\vartheta_{k}$. Covariate effects are included in the model through the parameters $\vartheta(x)$. By defining the regression parameters as a non-decreasing function of the covariates, we establish their relationship with the covariates. The regression parameter estimates are obtained by numerically maximizing the log-likelihood function. For this, we use the `nlminb' function in the R-package.

Cure fraction is calculated as the limit of survival function in defective models. The overall survival function is $S(t,x)=\prod_{j=1}^{k}S_{j}(t,x)$. Then the cure fraction is given by
\begin{equation}
    p(x)=\lim_{t\rightarrow \infty}S(t;x)=\lim_{t\rightarrow \infty}\left\{\prod_{j=1}^{k}S_j(t;.x)\right\}=\prod_{j=1}^{k}p_{j}(x).
\end{equation}
The asymptotic properties of model parameters follows from the asymptotic properties of the maximum likelihood estimators. Under certain regularity conditions, $\widehat{\vartheta}$ has
an asymptotic multivariate normal distribution with mean $\vartheta$ and variance and covariance matrix $\Sigma (\widehat{\vartheta})$, which is estimated by
\begin{equation*}
    \widehat{\Sigma} (\widehat{\vartheta}) = \left \{ \frac{-\partial^2 l(\vartheta ; D)}{\partial \vartheta \partial \vartheta^{T} } 
\left.\begin{matrix}
\\ 
\end{matrix}\right|_{\vartheta= \widehat\vartheta }\right  \}^{-1}.
\end{equation*}
An approximate $100(1-\alpha)\%$ confidence interval for $\vartheta_{j},j=1,2, \dots, k$ is $ (\widehat{\vartheta _{j}}- Z_{\alpha/2}\sqrt{\widehat{\Sigma}^{jj}},\widehat{\vartheta _{j}}+ Z_{\alpha/2}\sqrt{\widehat{\Sigma}^{jj}})$, where $\widehat{\Sigma}^{jj}$ denote the estimate of the  $j$-th diagonal element of the inverse of variance and covariance matrix and $Z_{\alpha}$ denotes the $100(1-\alpha)$ percentile of the standard normal random variable.

\subsection{Gompertz Defective Model}
The Gompertz distribution is used in many real-life applications, including medical and actuarial studies. One important characteristic of the Gompertz distribution is its exponentially increasing failure rate, making it suitable for modeling highly negatively skewed data in survival analysis. The survival function of the Gompertz model is given by
\begin{equation}
    S(t;\vartheta) =e^{-\frac{b}{a}(e^{at}-1)},
\end{equation}
with shape parameter $a>0$, scale parameter $b>0$, $t>0$, and $\vartheta=(a,b)^{'}$. When the shape parameter $a$ is negative, the Gompertz distribution becomes an improper distribution and the cure fraction is calculated as the limit of the survival function as follows:
\begin{equation*}
    \lim_{t\rightarrow \infty} S(t;\vartheta) = \lim_{t\rightarrow \infty} e^{-\frac{b}{a}(e^{at}-1)} = e^{\frac{b}{a}} = p \hspace{0.25cm} \in (0,1).
\end{equation*}
We propose a defective Gompertz distribution to model the cause specific survival function due to each cause under the assumption that $k$ causes are independent.
 It is given by
\begin{equation}
   S_{j}(t;\vartheta_{j}) =e^{-\frac{b_{j}}{a_{j}}(e^{a_{j}t}-1)},
\end{equation}
with $\vartheta_{j}=(a_{j},b_{j})$, where $b_{j}>0$ and $a_{j}<0$.
The covariate effects are included in the model through the parameters $a_{j}$ and $b_{j}$. The scale parameter $b_{j}$ is related to the covariate $x$ by the log-link function,
\begin{equation*}
    b_j(x) = \exp(x^{'}\beta_j),~~~~  j=1,2,\dots k,
\end{equation*}
where $x^{'} = (1,x_{1},x_{2}, \ldots,x_{p})$ and $\beta_j^{'}= (\beta_{0j},\beta_{1j}, \ldots, \beta_{pj})$ are the set of covariates and regression coefficients, respectively. 
We link the shape parameter to the covariate $x$ using an identity link function, allowing it to take both negative and positive values. It is given by
\begin{equation*}
    a_j(x) = x^{'}\gamma_j, ~~~~~  j=1,2,\dots k,
\end{equation*}
where $x^{'} = (1,x_{1},x_{2}, \ldots,x_{p})$ and $\gamma_j^{'}=(\gamma_{0j},\gamma_{1j},\ldots,\gamma_{pj})$  are the set of covariates and the regression coefficients, respectively, for each cause. A negative value of the shape parameter $a_j(x) = x^{'}\gamma_j$ for $j=1,2,\dots k$, ensures the presence of cure proportion in the population. 

The overall likelihood function for the defective Gompertz distribution is given by
\small
\begin{align}
     L(\vartheta )= & \prod_{i=1}^{n} \Bigg \{ \prod_{j=1}^{k}\left \{ \exp\left \{ -\frac{exp(x^{'}\beta_{j})}{x^{'}\gamma_{j}} (e^{x^{'}\gamma_{j}U_{i}}-1)\right \}-\exp\left \{ -\frac{exp(x^{'}\beta_{j})}{x^{'}\gamma_{j}} (e^{x^{'}\gamma_{j}V_{i}}-1)\right \} \right \}^{\delta _{ij}} \\ \nonumber & \left \{\prod_{j=1}^{k}\exp\left \{ -\frac{exp(x^{'}\beta_{j})}{x^{'}\gamma_{j}} (e^{x^{'}\gamma_{j}U_{i}}-1)\right \} \right \}^{\delta_{i} } \Bigg \}.
\end{align}
\normalsize
The corresponding log-likelihood function is given by
\small
\begin{align}
    l(\vartheta)= &\sum_{i=1}^{n}  \Bigg \{ \sum_{j=1}^{k}{\delta _{ij}}\log \left \{ \exp\left \{ -\frac{exp(x^{'}\beta_{j})}{x^{'}\gamma_{j}} (e^{x^{'}\gamma_{j}U_{i}}-1)\right \}-\exp\left \{ -\frac{exp(x^{'}\beta_{j})}{x^{'}\gamma_{j}} (e^{x^{'}\gamma_{j}V_{i}}-1)\right \} \right \} \\ \nonumber & +{\delta_{i} }\sum_{j=1}^{k} \left \{ -\frac{exp(x^{'}\beta_{j})}{x^{'}\gamma_{j}} (e^{x^{'}\gamma_{j}U_{i}}-1) \right \} \Bigg \}.
\end{align}
\normalsize
Since the log-likelihood function does not have a closed form, we use numerical methods for maximizing the log-likelihood function and to obtain estimates of the regression parameters. We use the `nlminb' function in R package for this purpose. Maximizing the log-likelihood function give the estimators $\hat a_j(x)$ and $\hat b_j(x)$ for $j=1,2,\dots k.$ The overall cure fraction is then calculated as
\begin{equation}
   \hat p(x)= \prod_{j=1}^{k}e^{\frac{\hat b_j(x)}{\hat a_j(x)}}.
\end{equation}

\subsection{Inverse Gaussian Defective Model}
The inverse Gaussian distribution is derived as the probability distribution of the first passage time for one-dimensional Brownian motion, or the Wiener process. This interpretation of the inverse Gaussian random variable as a first passage time suggests its potential applications in studying lifetimes or event occurrences across various fields. The survival function of the inverse Gaussian model is given by
\begin{equation}
    S(t;\vartheta)=1-\left [ \Phi \left ( \frac{-1+at}{\sqrt{bt}} \right ) + e^{2a/b}\Phi \left ( \frac{-1-at}{\sqrt{bt}} \right ) \right ],
\end{equation}
with shape parameter $a>0$, scale parameter $b>0$, $t>0$, and $\vartheta=(a,b)^{'}$. When the shape parameter $a$ is negative, the inverse Gaussian distribution becomes defective. The cure fraction is given by
\begin{equation*}
    \lim_{t\rightarrow \infty} S(t;\vartheta) = \lim_{t\rightarrow \infty}1-\left [ \Phi \left ( \frac{-1+at}{\sqrt{bt}} \right ) + e^{2a/b}\Phi \left ( \frac{-1-at}{\sqrt{bt}} \right ) \right ] = 1-e^{\frac{2a}{b}} = p \hspace{0.25cm} \in (0,1).
\end{equation*}

We assume that the cause specific survival function due to each cause follows a defective inverse Gaussian distribution under the assumption that $k$ causes are independent. It is given by
\begin{equation}
    S_{j}(t;\vartheta_{j})=1-\left [ \Phi \left ( \frac{-1+a_{j}t}{\sqrt{b_{j}t}} \right ) + e^{2a_{j}/b_{j}}\Phi \left ( \frac{-1-a_{j}t}{\sqrt{b_{j}t}} \right ) \right ],
\end{equation} 
with $\vartheta_{j}=(a_{j},b_{j})$, where $b_{j}>0$ and $a_{j}<0$.

Similar to the Gompertz defective model, the covariate effects are incorporated to the proposed model through the model parameters $a_{j}$ and $b_{j}$. The scale parameter $b_{j}>0$ is linked to the covariate $x$ by using the log-link function, while the shape parameter $a_j \in \mathbb{R}$ and the covariate $x$ is modeled using the identity link function as follows:
\begin{equation*}
    b_j(x) = \exp(x^{'}\beta_j),~~~~  j=1,2,\dots, k
\end{equation*}
and
\begin{equation*}
    a_j(x) = x^{'}\gamma_j, ~~~~~  j=1,2,\dots ,k,
\end{equation*}
where $x^{'} = (1,x_{1},x_{2}, \ldots,x_{p})$,  $\beta_j^{'}= (\beta_{0j},\beta_{1j}, \ldots, \beta_{pj})$, and $\gamma_j^{'}=(\gamma_{0j},\gamma_{1j},\ldots,\gamma_{pj})$  are the set of covariates and the regression coefficients, respectively, for each cause. To confirm the presence of cure proportion in the population,  $a_j(x) = x^{'}\gamma_j$ for $j=1,2,\dots k$, need to be negative.  

The overall likelihood function for the inverse Gaussian distribution is given by
\begin{align}
    L(\vartheta)= & \prod_{i=1}^{n}\Bigg\{\prod_{j=1}^{k}\Bigg\{ \Phi\left(\frac{-1+x^{'}\vartheta_{j}V_{i}}{\sqrt{e^{x^{'}\beta_{j}}V_{i}}}  \right)-\Phi\left(\frac{-1+x^{'}\vartheta_{j}U_{i}}{\sqrt{e^{x^{'}\beta_{j}}U_{i}}}  \right) \\ \nonumber & +e^{\frac{2x^{'}\vartheta_{j}}{e^{x^{'}\beta_{j}}}} \left\{  \Phi\left(\frac{-1-x^{'}\vartheta_{j}V_{i}}{\sqrt{e^{x^{'}\beta_{j}}V_{i}}}\right)-\Phi\left(\frac{-1-x^{'}\vartheta_{j}U_{i}}{\sqrt{e^{x^{'}\beta_{j}}U_{i}}}\right)\right\} \Bigg\}^{\delta_{ij}} \\ \nonumber & \prod_{j=1}^{k}\left\{   1- \Phi\left(\frac{-1+x^{'}\vartheta_{j}U_{i}}{\sqrt{e^{x^{'}\beta_{j}}U_{i}}}  \right) -e^{\frac{2x^{'}\vartheta_{j}}{e^{x^{'}\beta_{j}}}}\Phi\left(\frac{-1-x^{'}\vartheta_{j}U_{i}}{\sqrt{e^{x^{'}\beta_{j}}U_{i}}}\right)\right\}^{\delta_{i}}\Bigg\}.
\end{align}

The corresponding log-likelihood function is given by
 \begin{align}
     l(\vartheta)= & \sum_{i=1}^{n}\Bigg\{\sum_{j=1}^{k}\delta_{ij}\log \Bigg\{ \Phi\left(\frac{-1+x^{'}\vartheta_{j}V_{i}}{\sqrt{e^{x^{'}\beta_{j}}V_{i}}}  \right)-\Phi\left(\frac{-1+x^{'}\vartheta_{j}U_{i}}{\sqrt{e^{x^{'}\beta_{j}}U_{i}}}  \right)   \\ \nonumber & +e^{\frac{2x^{'}\vartheta_{j}}{e^{x^{'}\beta_{j}}}} \left\{  \Phi\left(\frac{-1-x^{'}\vartheta_{j}V_{i}}{\sqrt{e^{x^{'}\beta_{j}}V_{i}}}\right)-\Phi\left(\frac{-1-x^{'}\vartheta_{j}U_{i}}{\sqrt{e^{x^{'}\beta_{j}}U_{i}}}\right)\right\} \Bigg\}  \\ \nonumber & + \delta_{i}\sum_{j=1}^{k}\log\left\{   1- \Phi\left(\frac{-1+x^{'}\vartheta_{j}U_{i}}{\sqrt{e^{x^{'}\beta_{j}}U_{i}}}  \right) -e^{\frac{2x^{'}\vartheta_{j}}{e^{x^{'}\beta_{j}}}}\Phi\left(\frac{-1-x^{'}\vartheta_{j}U_{i}}{\sqrt{e^{x^{'}\beta_{j}}U_{i}}}\right)\right\}\Bigg\}.
 \end{align}
The maximum likelihood estimates are obtained by numerically maximizing the log-likelihood function. We use the `nlminb` function in R to compute these estimates. Let $\widehat{a}_{j}$ and $\widehat{b}_{j}$ denote the MLE of $a_{j}$ and $b_{j}$, respectively. The overall cure fraction is estimated as
\begin{equation}
    \hat{p}(x)= \Big(\prod_{j=1}^{k}1-e^{\frac{2\hat{a}_j(x)}{\hat{b}_j(x)}}\Big).
\end{equation}

\section{Simulation Studies}
We conduct a simulation study with $k=2$ causes of failure to assess the performance of the proposed estimation procedures. 
We consider two independent covariates $X_1$ and $X_2$, where $X_1$ is a Bernoulli random variable with $P(X_1=1)=0.5$ and $X_2 \sim \text{Uniform}(0,1)$. With the two covariates, the model we consider here is $a_j(x) = \gamma_{0j}+\gamma_{1j}x_{1} +\gamma_{2j}x_{2}$ and $b_j(x) = \exp(\beta_{0j}+\beta_{1j}x_{1} +\beta_{2j}x_{2})$, where $j=1,2$ indicating the causes of failure. The cure probabilities are given by $p_j=\lim_{t\rightarrow \infty }S_{j}(t)$, $j=1,2$, assuming causes of failure are independent. To ensure the presence of cured individuals, we generate lifetimes greater than $V_{i}$ from an exponential distribution with rate parameter $\tau$, which is chosen randomly.
Interval censored competing risks data with two causes of failure is generated using the following algorithm:

\begin{enumerate}
    \item Determine the initial values $\gamma_{0j},$ $\gamma_{1j},$ $\gamma_{2j},$ $\beta_{0j},$ $\beta_{1j},$ and $\beta_{2j}$; $j=1,2$.
    \item For the $i$th subject, draw $X_{1i} \sim \text{Bernoulli}(0.5)\hspace{0.15cm} \text{and}\hspace{0.15cm} X_{2i} \sim \text{Uniform}(0,1)$.
    \item Generate $M_i \sim$ Bernoulli $(1-p)$, where $p=p_{1}*p_{2}$.
    \item If $M_i=0$, set $T_i^{'}= \infty$. If $M_i=1$, generate $T_{i}$ using the inverse probability transformation method. 
    \item Generate $U_i^{'} \sim $ Uniform $(0,\max(T_i^{'}))$, considering only the finite $T_i^{'}$.
    \item Let $T_i = \min(T_i^{'},U_i^{'})$.
     \item If $\min (T_i^{'},U_i^{'})=U_{i}^{'}$, set $U_{i}=U_{i}^{'}$ and $V_{i}=\infty$.
    \item If $\min (T_i^{'},U_i^{'})=T_{i}^{'}$, generate $\textit{len}_{i}$ form Uniform(0.2,0.7) and $\textit{l}_{i}$ from Uniform(0,1). Then create intervals $(0,l_i],(l_{i},l_{i}+len_{i}], \ldots, (l_{i}+t \times len_{i},\infty), t=1,2, \ldots,$ and choose $(U_{i},V_{i})$ that satisfies $U_{i}<T_{i}^{'}\leq V_{i}$.
    \item The causes of failure are randomly generated by a binomial distribution with cell probabilities $1-p_{j}$ for $j=1,2$, where $p_{j}$ is the cure probability of the corresponding defective model.
    \item Generate $J_i= I(T_i^{'} \le U_i^{'})*$Cause.  If $J_i =1$, set the event occurred due to cause $1$, if $J_i =2$, set the event occurred due to cause $2$. Otherwise the event is considered censored.
    \item The dataset for the $i$-{th} subject is $(U_{i},V_{i},J_{i},X_{1i},X_{2i})$, $i=1,2,\ldots, n$.
\end{enumerate}
 The process is repeated 2000 times for different sample sizes $n=100$, $200$ and $500$. For each sample size, mean square error (MSE), absolute bias, and 90\% and 95\% coverage probability (CP) are calculated.

In the simulation studies, we consider two covariates: a discrete covariate $X_1$ and a continuous covariate $X_{2}$. The discrete covariate $X_1$ takes the values 0 and 1. For notional convenience, we categorize the individuals with $X_{1}=0$  into group 1 and with $X_{1}=1$ into group 2. The continuous covariate $X_2$ is generated from a uniform distribution in the interval $(0,1)$. If the generated value of $X_2$ is less than the median value, we assign the individual to group 3; otherwise we assign them to group 4. We denote $p_{13}$ as the cure rate of individuals belonging to groups 1 and 3, and $p_{14}$ as the cure rate of individuals belonging to groups 1 and 4. Similarly, we denote $p_{23}$ as the cure rate of individuals belonging to groups 2 and 3, and $p_{24}$ as the cure rate of individuals belonging to groups 2 and 4. 

To calculate the absolute bias, MSE, and CP of the model parameters for Gompertz defective model,
the initial values chosen are 
($\gamma_{01}$, $\gamma_{11}$, $\gamma_{21}$, $\gamma_{02}$, $\gamma_{12}$, $\gamma_{22}$)= (-0.2,-0.4,-0.6,-0.2,-0.5,-0.7) and ($\beta_{01}$, $\beta_{11}$, $\beta_{21}$, $\beta_{02}$, $\beta_{12}$, $\beta_{22}$)= (-2,1,1,1.5,-2,1,2). The initial values of the cure rates $p_{13}$, $p_{14}$, $p_{23}$, and $p_{24}$ calculated by using the equation
\begin{equation*}
    p(x)= \prod_{j=1}^{2}e^{\frac{ b_j(x)}{ a_j(x)}},
\end{equation*}
are 0.25, 0.15, 0.32, and 0.03, respectively.
Table \ref{tab:simgom} gives the absolute bias, MSE, and CP of the model parameters. From Table \ref{tab:simgom}, we observe that as the sample size increases, both absolute bias and MSE decrease, and the coverage probability remains close to the chosen level. Table \ref{tab:gocure} presents the absolute bias and MSE of the cure rate for different categories. From Table \ref{tab:gocure}, it is clear that as the sample size increases, absolute bias and MSE decrease. 

For the inverse Gaussian defective model, the initial values we consider are ($\gamma_{01}$, $\gamma_{11}$, $\gamma_{21}$, $\gamma_{02}$, $\gamma_{12}$, $\gamma_{22}$)= (-0.1,-0.3,-0.5,-0.2,-0.4,-0.6) and ($\beta_{01}$, $\beta_{11}$, $\beta_{21}$, $\beta_{02}$, $\beta_{12}$, $\beta_{22}$)= (-1.5,1,2,-1,1,2). Then we calculate initial values of the cure probability using the equation
\begin{equation*}
    p(x)=\prod_{j=1}^{2} \Big (  1-e^{2 a_j(x)/  b_j(x)}\Big).
\end{equation*}
The initial values of the cure rate for categories $p_{13}$, $p_{14}$, $p_{23}$, and $p_{24}$, obtained using the equation above, are 0.39, 0.23, 0.52, and 0.09, respectively. The absolute bias, MSE, and CP of the model parameters using an inverse Gaussian defective model are presented in Table \ref{tab:igsim}. Table \ref{tab:igcure} shows the absolute bias and MSE of the cure rate for different categories. From both tables, it is evident that as the sample size increases, absolute bias and MSE decrease. Additionally, from Table \ref{tab:igsim}, we observe that the estimated CP approaches the chosen level of significance. Thus, the simulation results of both models confirm their effectiveness.

\begin{table}[h]
\caption{Absolute bias, MSE, and 90\% and 95\% CP of the regression parameters in Gompertz defective model}
\label{tab:simgom}
\scalebox{0.73}{
\begin{tabular}{cccccccccccccc}
\hline
 
                     &           & \multicolumn{12}{c}{$(\gamma_{01},\gamma_{11},\gamma_{21},\gamma_{02},\gamma_{12},\gamma_{22},\beta_{01},\beta_{11},\beta_{21},\beta_{02},\beta_{12},\beta_{22})$=(-0.2,-0.4,-0.6,-0.2,-0.5,-0.7,-2,1,1.5,-2,1,2)} \\ \cline{3-14} 
n                    &           & $\gamma_{01}$    & $\gamma_{11}$    & $\gamma_{21}$    & $\gamma_{02}$   & $\gamma_{12}$   & $\gamma_{22}$   & $\beta_{01}$   & $\beta_{11}$   & $\beta_{21}$   & $\beta_{02}$   & $\beta_{12}$   & $\beta_{22}$   \\ \hline
\multirow{4}{*}{100} & MSE       & 0.0666           & 0.2519           & 0.3381           & 0.0688          & 0.2348          & 0.2979          & 0.2983         & 0.2341         & 0.6157         & 0.2684         & 0.2033         & 0.4124         \\
                     & BIAS      & 0.0712           & 0.1967           & 0.2045           & 0.0743          & 0.1155          & 0.0561          & 0.3238         & 0.1621         & 0.4205         & 0.2795         & 0.1268         & 0.0809         \\
                     & CP (90\%) & 0.9124           & 0.8083           & 0.8472           & 0.9086          & 0.8914          & 0.8924          & 0.7293         & 0.8273         & 0.7994         & 0.7364         & 0.8775         & 0.9195         \\
                     & CP (95\%) & 0.9452           & 0.8794           & 0.9038           & 0.9445          & 0.9417          & 0.9519          & 0.8472         & 0.9094         & 0.8621         & 0.8325         & 0.939          & 0.9572         \\ \hline
\multirow{4}{*}{300} & MSE       & 0.0138           & 0.0699           & 0.1135           & 0.0133          & 0.0493          & 0.0737          & 0.1597         & 0.0783         & 0.3036         & 0.1497         & 0.067          & 0.1347         \\
                     & BIAS      & 0.0371           & 0.1449           & 0.1772           & 0.0372          & 0.0544          & 0.0579          & 0.3185         & 0.1253         & 0.3951         & 0.2959         & 0.0829         & 0.0309         \\
                     & CP (90\%) & 0.9217           & 0.8564           & 0.8993           & 0.9197          & 0.9074          & 0.9043          & 0.8044         & 0.8691         & 0.8664         & 0.7994         & 0.8923         & 0.9324         \\
                     & CP (95\%) & 0.9556           & 0.9182           & 0.9518           & 0.9584          & 0.9562          & 0.9591          & 0.8746         & 0.9361         & 0.9261         & 0.8921         & 0.9419         & 0.9721         \\ \hline
\multirow{4}{*}{500} & MSE       & 0.0077           & 0.0462           & 0.0682           & 0.0065          & 0.0262          & 0.0387          & 0.1384         & 0.0544         & 0.2167         & 0.1215         & 0.0391         & 0.0759         \\
                     & BIAS      & 0.0345           & 0.1427           & 0.1599           & 0.0267          & 0.0337          & 0.0497          & 0.3064         & 0.1206         & 0.3627         & 0.2885         & 0.0841         & 0.0268         \\
                     & CP (90\%) & 0.9264           & 0.9247           & 0.9258           & 0.9314          & 0.9275          & 0.9186          & 0.8687         & 0.8942         & 0.8952         & 0.8932         & 0.8961         & 0.9428         \\
                     & CP (95\%) & 0.9692           & 0.9774           & 0.9754           & 0.9617          & 0.9658          & 0.9634          & 0.9271         & 0.9526         & 0.9447         & 0.9374         & 0.9453         & 0.9753         \\ \hline
\end{tabular}
}
\end{table}

\begin{table}[h]
\centering
\caption{Absolute bias and MSE of cure rate of simulated samples of the Gompertz defective model}
\label{tab:gocure}
\begin{tabular}{cccccc}
\hline
\multicolumn{6}{c}{$(p_{13},p_{14},p_{23},p_{24})$ = (0.25,0.15,0.32,0.03)} \\ \hline
n                    &      & $p_{13}$ & $p_{14}$ & $p_{23}$ & $p_{24}$ \\ \hline
\multirow{2}{*}{100} & Bias & 0.0698    & 0.0806    & 0.0903    & 0.0291    \\
                     & MSE  & 0.0055    & 0.0021    & 0.0039    & 0.0005    \\ \hline
\multirow{2}{*}{300} & Bias & 0.0635    & 0.0796    & 0.0873    & 0.0287    \\
                     & MSE  & 0.0031    & 0.0011    & 0.0022    & 0.0002    \\ \hline
\multirow{2}{*}{500} & Bias & 0.0419    & 0.0787    & 0.0835    & 0.0254    \\
                     & MSE  & 0.0024    & 0.0008    & 0.0017    & 0.0001    \\ \hline
\end{tabular}
\end{table}

\begin{table}[h]
\caption{Absolute bias, MSE, and 90\% and 95\% CP of the regression parameters in inverse Gaussian defective model}
\label{tab:igsim}
\scalebox{0.73}{
\begin{tabular}{cccccccccccccc}
\hline
&           & \multicolumn{12}{c}{$(\gamma_{01},\gamma_{11},\gamma_{21},\gamma_{02},\gamma_{12},\gamma_{22},\beta_{01},\beta_{11},\beta_{21},\beta_{02},\beta_{12},\beta_{22})$=(-0.1,-0.3,-0.5,-0.2,-0.4,-0.6,-1.5,1,2,-1,1,2)} \\ \cline{3-14} 
n                    &           & $\gamma_{01}$    & $\gamma_{11}$    & $\gamma_{21}$    & $\gamma_{02}$   & $\gamma_{12}$   & $\gamma_{22}$   & $\beta_{01}$   & $\beta_{11}$   & $\beta_{21}$   & $\beta_{02}$   & $\beta_{12}$   & $\beta_{22}$   \\ \hline
\multirow{4}{*}{100} & MSE       & 0.1193           & 0.3789           & 0.5358           & 0.1519          & 0.4002          & 0.6506          & 0.3219         & 0.3532         & 0.7694         & 0.4097         & 0.3801         & 0.8312         \\
                     & BIAS      & 0.1116           & 0.1413           & 0.0647           & 0.0589          & 0.0813          & 0.0387          & 0.2438         & 0.1995         & 0.3717         & 0.2603         & 0.1728         & 0.2354         \\
                     & CP (90\%) & 0.8439           & 0.8549           & 0.8610           & 0.8163          & 0.8338          & 0.8651          & 0.8392         & 0.8533         & 0.8422         & 0.8283         & 0.8654         & 0.8691         \\
                     & CP (95\%) & 0.8710           & 0.8932           & 0.9023           & 0.8640          & 0.8811          & 0.9053          & 0.9134         & 0.9162         & 0.9114         & 0.8941         & 0.9241         & 0.9209         \\ \hline
\multirow{4}{*}{300} & MSE       & 0.0242           & 0.0816           & 0.1368           & 0.0289          & 0.0721          & 0.1403          & 0.1439         & 0.1194         & 0.3113         & 0.1563         & 0.1145         & 0.2763         \\
                     & BIAS      & 0.0734           & 0.0982           & 0.0597           & 0.0462          & 0.0077          & 0.0248          & 0.2364         & 0.1622         & 0.3296         & 0.2338         & 0.1326         & 0.2315         \\
                     & CP (90\%) & 0.9391           & 0.9125           & 0.9214           & 0.8194          & 0.9041          & 0.9113          & 0.8981         & 0.8874         & 0.8973         & 0.8794         & 0.8841         & 0.8972         \\
                     & CP (95\%) & 0.9673           & 0.9673           & 0.9653           & 0.8772          & 0.9534          & 0.9592          & 0.9432         & 0.9241         & 0.9363         & 0.9293         & 0.9385         & 0.9464         \\ \hline
\multirow{4}{*}{500} & MSE       & 0.0157           & 0.0446           & 0.0731           & 0.0183          & 0.0412          & 0.0852          & 0.1123         & 0.0777         & 0.2203         & 0.1148         & 0.0745         & 0.1953         \\
                     & BIAS      & 0.0525           & 0.0766           & 0.0455           & 0.0186          & 0.0009          & 0.0128          & 0.1861         & 0.1506         & 0.3199         & 0.2318         & 0.1207         & 0.2295         \\
                     & CP (90\%) & 0.9481           & 0.9214           & 0.9321           & 0.8462          & 0.9164          & 0.9314          & 0.9184         & 0.8922         & 0.9164         & 0.8902         & 0.8961         & 0.9264         \\
                     & CP (95\%) & 0.9773           & 0.9733           & 0.9723           & 0.8964          & 0.9592          & 0.9682          & 0.9546         & 0.9436         & 0.9556         & 0.9365         & 0.9444         & 0.9637         \\ \cline{1-14} 
\end{tabular}
}
\end{table}

\begin{table}[h]
\centering
\caption{Absolute bias and MSE of cure rate of simulated samples of the inverse Gaussian defective model}
\label{tab:igcure}
\begin{tabular}{cccccc}
\hline
\multicolumn{6}{c}{($p_{13},p_{14},p_{23},p_{24}$)=(0.39,0.23,0.51,0.09)} \\ \hline
n                      &        & $p_{13}$  & $p_{14}$  & $p_{23}$  & $p_{24}$  \\ \hline
\multirow{2}{*}{100}   & BIAS   & 0.1251     & 0.0904     & 0.0355     & 0.0417     \\
                       & MSE    & 0.0126     & 0.0045     & 0.0097     & 0.0018     \\ \hline
\multirow{2}{*}{300}   & BIAS   & 0.0482     & 0.0801     & 0.0605     & 0.0411     \\
                       & MSE    & 0.0091     & 0.0021     & 0.0047     & 0.0007     \\ \hline
\multirow{2}{*}{500}   & BIAS   & 0.0337     & 0.0755     & 0.0579     & 0.0377     \\
                       & MSE    & 0.0064     & 0.0015     & 0.0035     & 0.0005     \\ \hline
\end{tabular}
\end{table}

\section{Data Analysis}
We apply our methods to the dataset `pseudo.hiv.long' from the R package `intccr'. This artificial dataset is simulated to resemble an HIV clinical trial on loss to care and mortality in sub-Saharan Africa, which was studied by \cite{bakoyannis2017semiparametric}. The original data were collected by the IeDEA-EA (East African International Epidemiology Databases to Evaluate AIDS) Consortium which includes HIV care and treatment programs in Kenya, Uganda, and Tanzania. The data consists of 3000 patients who initiated antiretroviral treatment (ART). For each patient, neither loss to care nor death times are exactly known. However, it is known that the event occurred between two observation times $U$, the last clinical examination before the event since ART initiation, and $V$, the first clinical examination time after the event. The age, male gender indicator, and CD4 cell count at ART initiation, respectively, are the covariates in the data. To reduce the complexity of the analysis, we consider the patients with age less than or equal to 51 who initiated ART 
in our study. We consider the male gender indicator and CD4 cell count as covariates in our study. The male gender indicator is a discrete covariate, with 1 representing male patients and 0 representing female patients. The CD4 cell count is a continuous covariate ranging from 57 to 1118.

The events observed during HIV care include loss to care and death while in care. Patients who are lost to care are considered as cause 1, and those who die while in care are considered as cause 2. The data we use here include 1486 patients who initiated ART at age 51 or younger. Out of these 1486 patients, 985 patients do not experience any events throughout the follow-up period. During the follow-up period, 449 patients are lost to care and 52 patients die while in HIV care. Since a large number of patients do not experience any of the events, it indicates the presence of cured individuals. We ensure the presence of cured individuals, by plotting survival curve estimates obtained using the Turnbull estimator. We use the `ic\_np' function in the R package `icenReg' to plot the survival curve. To plot the survival curves, we categorize the data into two groups based on CD4 cell counts: values less than 207 and values greater than or equal to 207, with 207 as the median value.
\begin{figure}[h]
    \centering
    \includegraphics[width=1\linewidth]{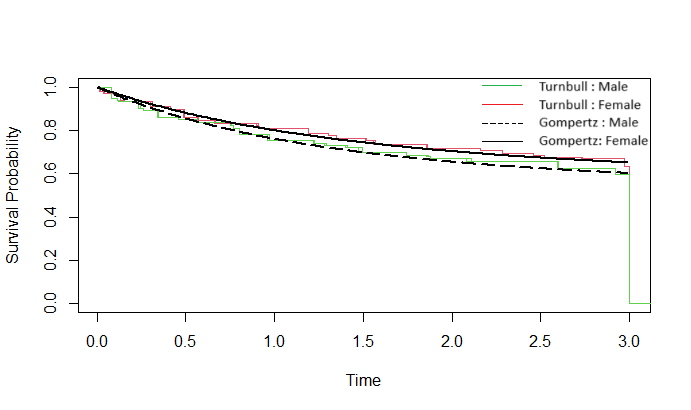}
    \caption{Survival curve obtained by Turnbull estimator and survival function estimates according to Gompertz model for the covariate male gender indicator}
    \label{gomale}
\end{figure}
\begin{figure}[h]
    \centering
    \includegraphics[width=1\linewidth]{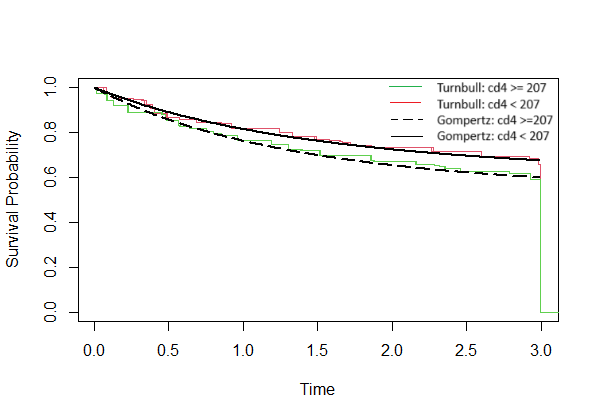}
    \caption{Survival curve obtained by Turnbull estimator and survival function estimates according to Gompertz model for the covariate CD4 cell count}
    \label{goCD4}
\end{figure}
\begin{figure}[h]
    \centering
    \includegraphics[width=1\linewidth]{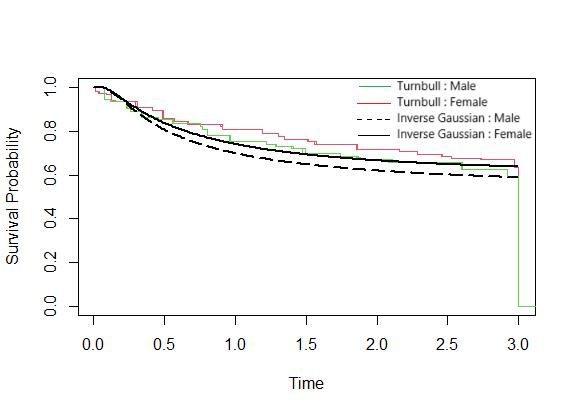}
    \caption{Survival curve obtained by Turnbull estimator and survival function estimates according to inverse Gaussian model for the covariate male gender indicator}
    \label{fig:igmale}
\end{figure}
\begin{figure}[h]
    \centering
    \includegraphics[width=1\linewidth]{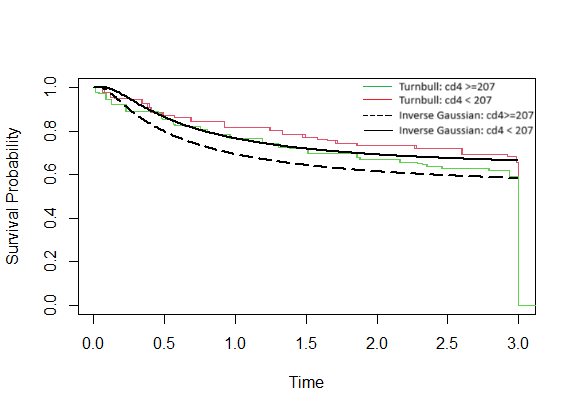}
    \caption{Survival curve obtained by Turnbull estimator and survival function estimates according to inverse Gaussian model for the covariate CD4 cell count}
    \label{fig:igCD4}
\end{figure}

Figures \ref{gomale} and \ref{goCD4} display the survival curve obtained by the Turnbull estimator and survival function estimates
according to the Gompertz model for the covariate male gender indicator and CD4 cell count, respectively.
For female patients, the survival curve does not come close to zero, and it settles down around the value of 0.7 of the survival probability. In the case of male patients, the curve suddenly drops beyond the minimum survival probability of around 0.6. This occurs because the largest recorded time is the observed lifetime of a male patient. When considering the CD4 cell count, patients with a CD4 cell count greater than or equal to 207, where 207 represents the median value, display a survival curve that falls below a minimum survival probability. This indicates that the largest recorded time is the observed lifetime with a CD4 cell count greater than 207, and the survival curve levels off around a survival probability of 0.7 when the cell count is less than 207.

 Similarly, Figure \ref{fig:igmale} and Figure \ref{fig:igCD4} depict the survival curves obtained using the Turnbull estimator, along with the survival function estimates of the inverse Gaussian model for the covariates male gender indicator and CD4 cell count, respectively. For the male gender indicator, the curve settles around a survival probability of 0.6 for males and 0.65 for females. A plateau is noted around survival probabilities of 0.65 for CD4 cell counts below the median value and 0.6 for those at or above the median value. The sudden fall in survival probability is due to the fact that the largest recorded time is an observed lifetime. Hence, the survival curves confirm the presence of cured individuals in the population.

 For the model selection criteria, we calculate the Akaike Information Criterion (AIC) and Bayesian Information Criterion (BIC) values of the HIV data for the Gompertz and inverse Gaussian models. The AIC and BIC are defined as AIC $= -2L + 2k$ and BIC $= -2L +k \log n,$ where $L$ represents the maximized log likelihood function value, $k$ is the number of parameters in the fitted model, and $n$ is the sample size. To obtain the effect of covariates individually, we consider the male gender indicator and CD4 cell count separately.
 When considering the male gender indicator as a covariate, the AIC values are 3863 for the Gompertz distribution and 4007 for the inverse Gaussian distribution, with BIC values of 3905 and 4049, respectively. When using the CD4 cell count as the covariate, the AIC values are 3855 for the Gompertz distribution and 3996 for the inverse Gaussian distribution, and the BIC values are 3897 and 4038, respectively. The lower AIC and BIC values of the Gompertz defective model suggest that it is more appropriate for the HIV data for both covariates. This is further supported by the survival curves, as the Gompertz model closely aligns with the Turnbull estimator for both covariates: the male gender indicator and CD4 cell count.
 
 We first estimate the cure fraction for the HIV data using the Gompertz defective model and the inverse Gaussian defective model, considering the male gender indicator as a covariate. The regression parameters estimated using the maximum likelihood method, along with their standard errors and confidence intervals, are presented in Tables \ref{tab:malecovgo} and \ref{tab:igmalecov}, respectively.  We then estimate the cure fraction based on the CD4 cell count categories using the Gompertz and inverse Gaussian defective models. The regression parameters estimated using the maximum likelihood method, along with their standard errors and confidence intervals for the covariate CD4 cell count, are
summarized in Tables \ref{tab:CD4covgo} and \ref{tab:igCD4cov}, respectively.

\begin{table}[h]
\caption{MLE, SE, 95\% CI for the Gompertz defective model using pseudo HIV data with the covariate male gender indicator}
\label{tab:malecovgo}
\scalebox{0.9}{
\begin{tabular}{cccccccccc}
\hline
\multirow{2}{*}{Parameter} & \multirow{2}{*}{MLE} & \multirow{2}{*}{SE} & \multicolumn{2}{c}{CI(95\%)} & \multirow{2}{*}{Parameter} & \multirow{2}{*}{MLE} & \multirow{2}{*}{SE} & \multicolumn{2}{c}{CI(95\%)} \\ \cline{4-5} \cline{9-10} 
                           &                      &                     & Lower         & Upper        &                            &                      &                     & Lower         & Upper        \\ \hline
$\gamma_{01}$              & -0.6097              & 0.0835              & -0.7735       & -0.4459      & $\gamma_{02}$              & -0.0706              & 0.2171              & -0.4962       & 0.3550       \\
$\gamma_{11}$              & 0.0143               & 0.1453              & -0.2706       & 0.2992       & $\gamma_{12}$              & -0.5241              & 0.4173              & -1.3421       & 0.2939       \\
$\beta_{01}$               & -1.3028              & 0.0964              & -1.4919       & -1.1138      & $\beta_{02}$               & -3.8824              & 0.3271              & -4.5234       & -3.2413      \\
$\beta_{11}$               & 0.1672               & 0.1672              & -0.1605       & 0.4950       & $\beta_{12}$               & 0.7063               & 0.5409              & -0.3537       & 1.7665       \\ \hline
\multicolumn{2}{c}{p (female )}                   & 0.543               &               &              &                            &                      &                     &               &              \\
\multicolumn{2}{c}{p(male)}                       & 0.478                &               &              &                            &                      &                     &               &              \\ \hline
max $l(\vartheta)$         &                      &1923                     &               &              & AIC                        & 3863                 &                     &               &              \\
                           &                      &                     &               &              & CAIC                       & 3913                 &                     &               &              \\
                           &                      &                     &               &              & BIC                        & 3905                 &                     &               &              \\ \hline
\end{tabular}
}
\end{table}

\begin{table}[h]
\centering
\caption{MLE, SE, 95\% CI for the inverse Gaussian defective model using pseudo HIV data with the covariate male gender indicator}
\label{tab:igmalecov}
\scalebox{0.9}{
\begin{tabular}{cccccccccc}
\hline
\multirow{2}{*}{Parameter} & \multirow{2}{*}{MLE} & \multirow{2}{*}{SE} & \multicolumn{2}{c}{CI(95\%)} & \multirow{2}{*}{Parameter} & \multirow{2}{*}{MLE} & \multirow{2}{*}{SE} & \multicolumn{2}{c}{CI(95\%)} \\ \cline{4-5} \cline{9-10} 
                           &                      &                     & Lower          & Upper       &                            &                      &                     & Lower          & Upper       \\ \hline
$\gamma_{01}$              & -0.8731              & 0.0770              & -1.0241        & -0.7221            & $\gamma_{02}$              & -0.6479              & 0.1055              & -0.8548        & -0.4410            \\
$\gamma_{11}$              & 0.1643               & 0.1304              & -0.0913        & 0.4201            & $\gamma_{12}$              & -0.7686              & 0.2867              & -1.3305        &-0.2066             \\
$\beta_{01}$               & 0.5277               & 0.0712              & 0.3881         & 0.6675            & $\beta_{02}$               & -0.7733              & 0.1438              & -1.0553        &  -0.4914            \\
$\beta_{11}$               & -0.0362              & 0.1276              & -0.2865        & 0.2140            & $\beta_{12}$               & 0.8154               & 0.2411              & 0.3427         &  1.2880           \\ \hline
\multicolumn{2}{c}{p (female )}                   & 0.604               &                &             &                            &                      &                     &                &             \\
\multicolumn{2}{c}{p(male)}                       & 0.541               &                &             &                            &                      &                     &                &             \\ \hline
max $l(\vartheta)$         &                 & 1995                    &                &             & AIC                        & 4007                 &                     &                &             \\
                           &                      &                     &                &             & CAIC                       & 4057                 &                     &                &             \\
                           &                      &                     &                &             & BIC                        & 4049                 &                     &                &             \\ \hline
\end{tabular}
}
\end{table}

\begin{table}[h!]
\centering
\caption{MLE, SE, 95\% CI for the Gompertz defective model using pseudo HIV data with the covariate CD4 cell count}
\label{tab:CD4covgo}
\scalebox{0.9}{
\begin{tabular}{cccccccccc}
\hline
\multirow{2}{*}{Parameter} & \multirow{2}{*}{MLE} & \multirow{2}{*}{SE} & \multicolumn{2}{c}{CI(95\%)} & \multirow{2}{*}{Parameter} & \multirow{2}{*}{MLE} & \multirow{2}{*}{SE} & \multicolumn{2}{c}{CI(95\%)} \\ \cline{4-5} \cline{9-10} 
                           &                      &                     & Lower         & Upper        &                            &                      &                     & Lower         & Upper        \\ \hline
$\gamma_{01}$              & -0.6225              & 0.1041              & -0.8267       & -0.4184      & $\gamma_{02}$              & -0.1251              & 0.2364              & -0.5885       & 0.3382       \\
$\gamma_{11}$              & 0.0376               & 0.1381              & -0.2331       & 0.3083       & $\gamma_{12}$              & -0.2451              & 0.3770              & -0.9841       & 0.4937       \\
$\beta_{01}$               & -1.4084              & 0.1204              & -1.6446       & -1.1723      & $\beta_{02}$               & -3.6971              & 0.3511              & -4.3854       & -3.0088      \\
$\beta_{11}$               & 0.2938               & 0.1593              & -0.0184       & 0.6061       & $\beta_{12}$               & 0.1034               & 0.5234              & -0.9225       & 1.1294       \\ \hline
\multicolumn{2}{c}{p (CD4 $\ge $ 207)}                   & 0.529               &               &              &                            &                      &                     &               &              \\
\multicolumn{2}{c}{p(CD4 $<$ 207)}                       & 0.553               &               &              &                            &                      &                     &               &              \\ \hline
max $l(\vartheta)$         &                      & 1919                &               &              & AIC                        & 3855                 &                     &               &              \\
                           &                      &                     &               &              & CAIC                       & 3905                 &                     &               &              \\
                           &                      &                     &               &              & BIC                        & 3897                 &                     &               &              \\ \hline
\end{tabular}
}
\end{table}

\begin{table}[tbh!]
\centering
\caption{MLE, SE, 95\% CI for the inverse Gaussian defective model using pseudo HIV data set with the covariate CD4 cell count}
\label{tab:igCD4cov}
\scalebox{0.9}{
\begin{tabular}{cccccccccc}
\hline
\multirow{2}{*}{Parameter} & \multirow{2}{*}{MLE} & \multirow{2}{*}{SE} & \multicolumn{2}{c}{CI(95\%)} & \multirow{2}{*}{Parameter} & \multirow{2}{*}{MLE} & \multirow{2}{*}{SE} & \multicolumn{2}{c}{CI(95\%)} \\ \cline{4-5} \cline{9-10} 
                           &                      &                     & Lower         & Upper        &                            &                      &                     & Lower         & Upper        \\ \hline
$\gamma_{01}$              & -0.7581              & 0.0857              & -0.9261       & -0.5901      & $\gamma_{02}$              & -0.9833              & 0.1421              & -1.2620       & -0.7046      \\
$\gamma_{11}$              & -0.0806              & 0.1235              & -0.3227       & 0.1615       & $\gamma_{12}$              & 0.2828               & 0.2081              & -0.1251       & 0.6908       \\
$\beta_{01}$               & 0.2934               & 0.0887              & 0.1194        & 0.4675       & $\beta_{02}$               & -0.3069              & 0.1412              & -0.5837       & -0.0301      \\
$\beta_{11}$               & 0.3875               & 0.1198              & 0.1527        & 0.6223       & $\beta_{12}$               & -0.4398              & 0.2410              & -0.9122       & 0.0325       \\ \hline
\multicolumn{2}{c}{p (CD4$\ge $207 )}             & 0.542               &               &              &                            &                      &                     &               &              \\
\multicolumn{2}{c}{p(CD4$<$207)}                  & 0.630               &               &              &                            &                      &                     &               &              \\ \hline
max $l(\vartheta)$         &                      & 1990                &               &              & AIC                        & 3996                 &                     &               &              \\
                           &                      &                     &               &              & CAIC                       & 4046                 &                     &               &              \\
                           &                      &                     &               &              & BIC                        & 4038                 &                     &               &              \\ \hline
\end{tabular}
}
\end{table}


 Table \ref{tab:malecovgo} presents the results for the male gender indicator using the Gompertz defective model, showing a cure rate of 0.478 for male patients and 0.543 for female patients. From Table \ref{tab:CD4covgo}, it is observed that, under the Gompertz defective model, patients with a CD4 cell count below 207 have a cure rate of 0.553, which is slightly higher than that of patients with a CD4 cell count greater than or equal to 207. Using the inverse Gaussian defective model, as shown in Table \ref{tab:igmalecov}, the cure rate for female patients is 0.604, while the cure rate for male patients is 0.478. Table \ref{tab:igCD4cov} shows that, for the inverse Gaussian defective model, patients with a CD4 cell count below 207 have a cure rate of 0.630, which is higher than the cure rate of 0.542 for those with a CD4 count of 207 or above. The survival curves confirm these results: for both the Gompertz and inverse Gaussian defective models, the survival curve for female patients is higher than that for male patients based on the male gender indicator. Similarly, for the CD4 cell count covariate, the survival curve for patients with a CD4 count below 207 lies above that for patients with a CD4 count of 207 or higher. The estimates derived from the inverse Gaussian defective model are higher than those obtained using the Gompertz defective model.

\section{Conclusions}\label{sec5}
     In this paper, we introduced two defective regression models viz. Gompertz and inverse Gaussian defective regression models for the analysis of interval-censored competing risks data in the presence of cured individuals. An advantage of the defective models is that they allow for the direct estimation of the cure rate by utilizing an improper distribution, whereas, in the mixture model approach, a specific parameter accounts for the cure rate. The defective models proposed here enable us to calculate the cure rate directly as $p=\prod_{j=1}^{k}S_{j}(t)$, where $S_{j}(t)$ is the survival function due to each cause $j$. 
     
     We conducted simulation studies to validate the performance of the proposed procedures in finite samples. The practical relevance and applicability of the proposed models were demonstrated using a real-life dataset. The AIC and BIC values of the dataset were calculated, and the model compatibility was assessed using the survival curve obtained with the Turnbull estimator. In this study, we consider the interval censored lifetimes, where various other forms of censoring may occur in lifetime studies.    
     Defective models that accommodate various forms of censoring in the presence of cured individuals are a topic of interest for future studies. 



\section*{Statements and Declarations}
\section*{Conflicts of Interest}
There are no conflicts of interest between the authors.
\section*{Data Availability}
The source of the data sets used for illustration purposes is mentioned in the manuscript.
\section*{Ethical Statement}
The submitted work is original and has not been published elsewhere. 
\section*{Funding}
The first author received funding from  Cochin University of Science and Technology to conduct the research.

\bibliographystyle{apalike}
\bibliography{refe}



\end{document}